\documentclass{ws-ijbc}
\usepackage{ws-rotating}     
\usepackage{graphicx}
\usepackage{epstopdf}
\begin{document}
\catchline{}{}{}{}{} 

\markboth{Anisha R. V. Kashyap and Kiran M. Kolwankar}{Single Element Nonlinear Chimney Model}

\title{Chaotic Properties of Single Element Nonlinear Chimney Model: Effect of Directionality}
\author{Anisha R. V. Kashyap}
\address{Department of Physics, University of Mumbai, Santa Cruz (E),\\
Mumbai 400 098, India\\Department of Physics, Ramniranjan Jhunjhunwala College, \\ Ghatkopar (W), Mumbai 400 086, India\\anisha.kashyap27@gmail.com}

\author{Kiran M. Kolwankar}
\address{Department of Physics, Ramniranjan Jhunjhunwala College, \\ Ghatkopar (W), Mumbai 400 086, India\\Kiran.Kolwankar@gmail.com}
\maketitle

\begin{history}
\received{(to be inserted by publisher)}
\end{history}

\begin{abstract}
We generalize the chimney model by introducing nonlinear restoring and gravitational forces for the purpose of modeling swaying of trees at high wind speeds. We have derived general equations governing the system using Lagrangian formulation. We have studied the simplest case of a single element in more detail. The governing equation we arrive at for this case has not been studied so far. We study the chaotic properties of this simple building block and also the effect of directionality in the wind on the chaotic properties. We also consider the special case of two elements.
\end{abstract}

\section{Introduction}

Though swaying of trees is cited as a standard example of a natural nonlinear system, surprisingly its complete nonlinear dynamical modelling has not yet been carried out in spite of its obvious applications. Dynamics of swaying of trees is of interest to forest scientists~\cite{Lan} as it has consequences to the losses occurred in stormy conditions. As a result, a typical question asked is how the shape of canopy determines its response to the wind.  Also, in computer animation~\cite{DRBR,AK,OK,HZXI} developing methods for realistically depicting the movement of trees is an active field. Here one requires the
appropriate dynamical equations modeling the motion to have better visual
effect.

In the past, various theoretical methods have been developed to describe the response of the tree to the wind load.
These include, considering a cantilever beam approximation, a partial differential equation for free vibrations of the beam~\cite{MM} or a chimney model consisting of coupled short oscillating sections~\cite{KG}. But none of these  incorporate the nonlinear restoring force and also the branched structure of a tree. However, there are some recent works which have begun to take into account the nonlinear effect~\cite{LM}. Also, very recently, Murphy and Rudnicki [2012] have evolved a way to incorporate branching structure and also the nonlinearity in the model. In another work, Thecke, et al. [2011]  have constructed a Y-shaped branched model in order to understand the structural stability for possible applications to mechanical designs. Though these works have initiated the incorporation of nonlinear effects in the modelling of swaying trees, a complete nonlinear analysis of the phenomenon is still lacking. There are some handful of investigations done to study the resonance behaviour of plant stem based on mass and nonlinear flexural stiffness distributions. However, there are still many aspects, especially the chaoticity, which remain to be explored.

On the experimental front, substantial work~\cite{Lan}  has been carried out to measure the motion of the trees, hence different methods are used to record displacement, acceleration and velocity of the plant with the help of optical target monitoring~\cite{HLP}, inclinometer~\cite{SFP} and image correlation from videos~\cite{BDH}. The objectives of the experiments have been diverse, from studying the effect of wind velocity to the influence of aerial architecture.

We have begun a program to carry out this modelling \emph{ab initio} and
plan to carry out comparisons of the results thus obtained with experimental data either already available or carried out for the purpose. This work is the first step in this direction which introduces and analyses the simplest model which arises as a natural evolution in this process. It is not intended to include biological inputs at this stage but only to study the nonlinear dynamical aspects of the model.

Several computer animation studies (see, for example,~\cite{OTFFMC}), in order to make the animation realistic, assume that the wind is turbulent and use $1/f^\beta$ noise as a driving force. Our work, in fact, explores another point of view, that is, the question how much of the irregular motion of the trees 
is due to nonlinear restoring forces leading to chaotic behavior. Hence we consider the wind to be laminar and use simple driving forces as explained later. 

The paper is organized as follows. In section~\ref{se:model}, we introduce and explain our model which includes the derivation of the Lagrangian governing the system. This is followed by the section explaining the numerical results, the study of Lyapunov exponents for different values of driving frequencies and the effect of different parameters of the model on the chaotic properties. Then we end by some concluding discussions.

\section{The Model}\label{se:model}

\begin{figure}
\includegraphics[width=10cm,scale=15]{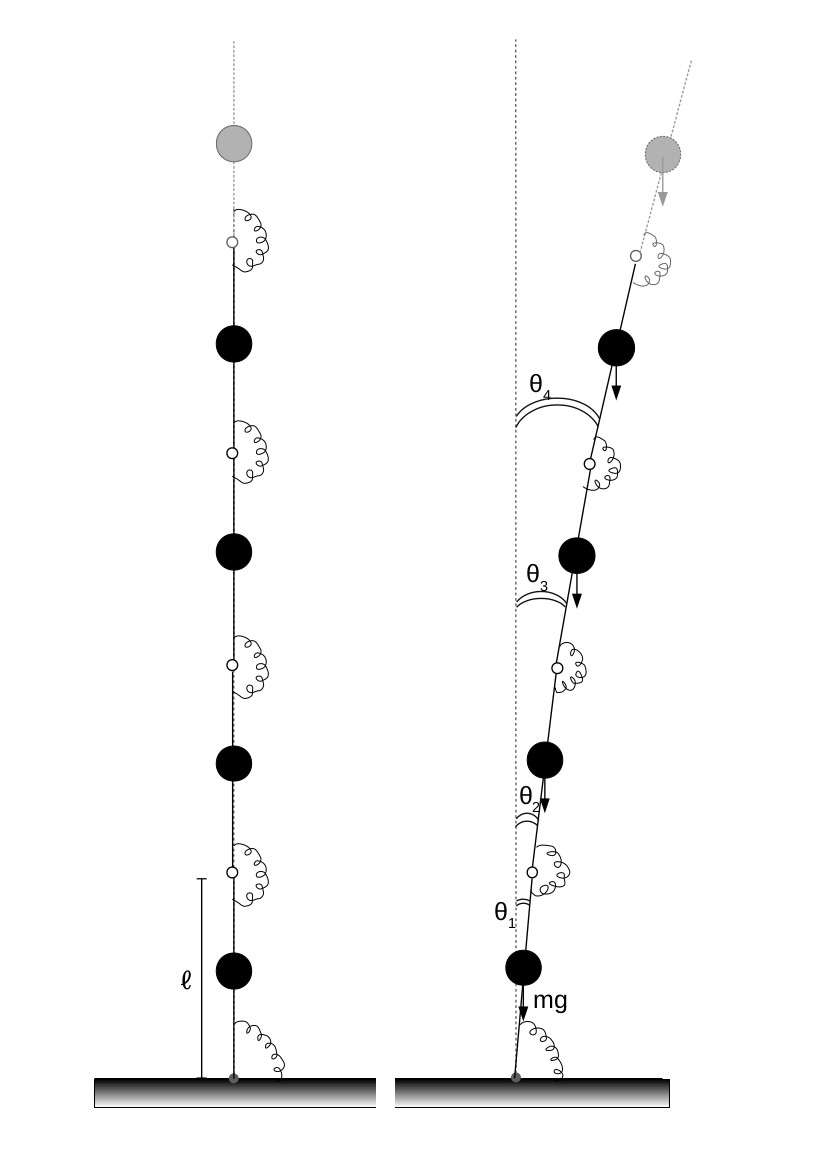}%
\caption{Schematic diagram of the Chimney model. Different segments are connected end to end with their mass concentrated at the center. There is a restoring force at the joints of the segments and also at the base of the lower most segment. There is a downward gravitational force acting on each segment.}\label{fig:chimney}
\end{figure}

Our starting point is the Chimney model which was studied in \cite{KG}.
As shown in Fig.~\ref{fig:chimney}, it consists of a vertical column made of several segments with a restoring force at the joints and a gravitational destabilising force. It has been used to understand the swaying motion of trees and hitherto formulated only using linear terms \cite{KG}.\footnote[1]{We however retain the word chimney in the name though the model may no longer be applicable to chimneys.} This choice of the model would allow us to easily add the branching structure at the later stage of the development.

\subsection{General formulation}
We have reformulated this problem using Lagrangian formulation and generalised to include nonlinearities in order to understand the motion of trees even at high wind velocities. As is clear from Fig.~\ref{fig:chimney}, the $\theta_i$ is the angle made by the $i^{th}$ element with the verticle and $m_i$ is the mass which is assumed to be concentrated at the center. Here we also assume that the lengths of all the elements are the same and equal to $\ell$. If there are $N$ number of elements and $(x_n,y_n)$ are the coordinates of the center of the $n^{th}$ ($1\leq n \leq N$) element, then we have
\( x_n = \sum_{i=1}^{n-1} \ell\sin\theta_i + \frac{\ell}{2} \sin\theta_n \)
and
\( y_n = \sum_{i=1}^{n-1} \ell\cos\theta_i + \frac{\ell}{2} \cos\theta_n. \)
Also, the components of velocities are given by
\( \dot{x}_n = \sum_{i=1}^{n-1} \ell\cos(\theta_i)\dot{\theta}_i + \frac{\ell}{2} \cos(\theta_n) \dot{\theta}_n \)
and
\( \dot{y}_n = -\sum_{i=1}^{n-1} \ell\sin(\theta_i)\dot{\theta}_i - \frac{\ell}{2} \sin(\theta_n) \dot{\theta}_n. \)

So the kinetic energy is given by 
\begin{eqnarray}\nonumber
T_N &=& \sum_{n=1}^{N} \frac{1}{2} m_n (\dot{x}_n^2 + \dot{y}_n^2 )\\
&=& \sum_{n=1}^{N} \frac{1}{2} m_n \left( \left(\sum_{i=1}^{n-1}\ell\cos(\theta_i)\dot{\theta}_i + \frac{\ell}{2} \cos(\theta_n) \dot{\theta}_n\right)\left(\sum_{j=1}^{n-1} \ell\cos(\theta_j)\dot{\theta}_j + \frac{\ell}{2} \cos(\theta_n) \dot{\theta}_n\right) \right.\nonumber \\
&& \;\;\;\;\;\; \left. +  \left(-\sum_{i=1}^{n-1}\ell\sin(\theta_i)\dot{\theta}_i - \frac{\ell}{2} \sin(\theta_n) \dot{\theta}_n\right)\left(-\sum_{j=1}^{n-1} \ell\sin(\theta_j)\dot{\theta}_j - \frac{\ell}{2} \sin(\theta_n) \dot{\theta}_n \right) \right) \nonumber
\end{eqnarray}
and after some simplification it becomes,
\begin{eqnarray}
T_N &=& \frac{\ell^2}{2}\sum_{n=1}^{N}m_n\left( \sum_{i=1}^{n-1}\sum_{j=1}^{n-1}\cos(\theta_i-\theta_j)\dot{\theta}_i\dot{\theta}_j+\sum_{i=1}^{n-1}\cos(\theta_n-\theta_i)\dot{\theta}_n\dot{\theta}_i+\frac{1}{4}{\dot{\theta}_n}^2\right) \nonumber
\end{eqnarray}
Now we interchange the sums and, after some algebra, obtain
\begin{eqnarray}
T_N&=& \ell^2\sum_{j=1}^{N}\sum_{i=j+1}^{N}\left(\Big(\cos(\theta_j-\theta_i)\Big)\dot{\theta}_j\dot{\theta}_i\bigg(\frac{m_i}{2}+\sum_{n=i+1}^{N}m_n\bigg)\right)\nonumber\\
&& \;\;\;\;\;\;+\frac{\ell^2}{2}\sum_{j=1}^{N}{\dot{\theta}_j}^2\left(  \frac{m_j}{4}+\sum_{i=j+1}^{N}m_i\right)
\end{eqnarray}
The potential energy is,
\begin{eqnarray}
V_N &=&\sum_{n=1}^{N}\left(m_ngy_n+\frac{1}{2}k_n(\theta_n-\theta_{n-1})^2+\frac{1}{4}k_n\alpha_n(\theta_n-\theta_{n-1})^4\right)\nonumber\\
&=&\sum_{n=1}^{N}\left(m_ng\left(\sum_{i=1}^{n-1} \ell\cos\theta_i + \frac{\ell}{2} \cos\theta_n\right)+\frac{1}{2}k_n(\theta_n-\theta_{n-1})^2+\frac{1}{4}k_n\alpha_n(\theta_n-\theta_{n-1})^4\right)\nonumber
\end{eqnarray}
and after rearranging terms it takes the form
\begin{eqnarray}\nonumber
V_N &=&g\ell\sum_{n=1}^{N}m_n\left(\sum_{i=1}^{n-1} \cos\theta_i + \frac{1}{2} \cos\theta_n\right)+\frac{1}{2}\sum_{n=1}^{N}k_n(\theta_n-\theta_{n-1})^2+\frac{1}{4}\sum_{n=1}^{N}k_n\alpha_n(\theta_n-\theta_{n-1})^4.
\end{eqnarray}
Now, again, interchanging the sums, we get
\begin{eqnarray}
V_N&=&g\ell\sum_{j=1}^{N}\cos\theta_j\bigg(\frac{m_j}{2}+\sum_{i=j+1}^{N}  m_i\bigg)+\frac{1}{2}\sum_{j=1}^{N}k_j(\theta_j-\theta_{j-1})^2+\frac{1}{4}\sum_{j=1}^{N}k_j\alpha_j(\theta_j-\theta_{j-1})^4\nonumber\\
\end{eqnarray}

Thus Lagrangian is written as:
\begin{eqnarray}\nonumber
L_N&=&T_N-V_N\\
&=&\frac{\ell^2}{2}\sum_{j=1}^{N}\dot{\theta_j}^2\bigg(\frac{m_j}{4}+\sum_{i=j+1}^{N}m_i\bigg)+\ell^2\sum_{j=1}^{N}\sum_{i=j+1}^{N}\dot{\theta_j}\dot{\theta_i}\cos(\theta_i-\theta_j)\bigg(\frac{m_i}{2}+\sum_{n=i+1}^{N}m_n\bigg)\nonumber\\
 && \;\;\; -g\ell\sum_{j=1}^{N}\cos\theta_j\bigg(\frac{m_j}{2}+\sum_{i=j+1}^{N}m_i\bigg)-\frac{1}{2}\sum_{j=1}^{N}k_j(\theta_j-\theta_{j-1})^2-\frac{1}{4}\sum_{j=1}^{N}k_j\alpha_j(\theta_j-\theta_{j-1})^4
\end{eqnarray}
If we write the cumulative mass of the segments above the segment $i$ as $M_i = \sum_{n=i+1}^{N}m_n$ then we get
\begin{eqnarray}
L_N
&=&\frac{\ell^2}{2}\sum_{j=1}^{N}\dot{\theta_j}^2\bigg(\frac{m_j}{4}+M_j\bigg)+\ell^2\sum_{j=1}^{N}\sum_{i=j+1}^{N}\dot{\theta_j}\dot{\theta_i}\cos(\theta_i-\theta_j)\bigg(\frac{m_i}{2}+M_i\bigg)\nonumber\\
 && \;\;\; -g\ell\sum_{j=1}^{N}\cos\theta_j\bigg(\frac{m_j}{2}+M_j\bigg)-\frac{1}{2}\sum_{j=1}^{N}k_j(\theta_j-\theta_{j-1})^2-\frac{1}{4}\sum_{j=1}^{N}k_j\alpha_j(\theta_j-\theta_{j-1})^4
\end{eqnarray}

\subsection{Special case of single element}
For the special case of just one segment, the Lagrangian takes the form:
\begin{eqnarray}
L_1&=&\frac{1}{2}\ell^2\Big(\frac{m_1}{4}\Big)\dot{\theta_1}^2-g\ell\Big(\frac{m_1}{2}\Big)\cos\theta_1-\frac{1}{2}k_1\theta_1^2-\frac{1}{4}k_1\alpha_1\theta_1^4
\end{eqnarray}
Thus Lagrangian equation of motion for single element (including nonlinear restoring force) is as follows
\begin{eqnarray}\nonumber
\frac{d}{dt}\bigg(\frac{\partial L_1}{\partial \dot{\theta_1}}\bigg)-\frac{\partial L_1}{\partial \theta_1}&=&Q_1\\
\frac{m_1\ell^2\ddot{\theta_1}}{4}-\frac{m_1g\ell\sin\theta_1}{2}+k_1\theta_1\big(1+\alpha_1\theta_{1}^2\big)&=&Q_1
\end{eqnarray}
where $Q_1$ is the total force in the direction of $\theta_1$ which does not arise from any potential. Here it consists of the dissipation force and the driving force.
To incorporate the dissipation,
the frictional force defined in terms of a function $\mathcal{F},$ known as Rayleigh's dissipation function, which is given by 
\begin{eqnarray}
\mathcal{F}&=&\frac{1}{2}\bigg(b_x{v_{x1}}^2+b_y{v_{y1}}^2\bigg)
\end{eqnarray}
Since the velocity components for single element are as follows:
\begin{eqnarray}
v_{x1}&=&\dot{x_1}=\frac{\ell}{2}\cos\theta_1\dot{\theta_1}\nonumber\\
v_{y1}&=&\dot{y_1}=-\frac{\ell}{2}\sin\theta_1\dot{\theta_1},\nonumber
\end{eqnarray}
the Rayleigh's dissipation function becomes
\begin{eqnarray}
\mathcal{F}&=&\frac{1}{2}\bigg(b_x\big({\frac{\ell}{2}\cos\theta_1\dot{\theta_1}}\big)^2+b_y\big({-\frac{\ell}{2}\sin\theta_1\dot{\theta_1}}\big)^2\bigg)
\end{eqnarray}
and with the assumption that the dissipation along each direction is the  same, i.e. $b_x=b_y=b$, it simplifies to
\begin{eqnarray}\nonumber
\mathcal{F}&=&\frac{1}{2}b\ell^2\Bigg(\frac{{\dot{\theta_1}}^2}{4}\Bigg)
\end{eqnarray}

As a result, the Lagrange's equation of motion for a single element system with the dissipative force, $\frac{\partial \mathcal{F}}{\partial \dot{\theta_1}}$, is given as
\begin{eqnarray}
\frac{d}{dt}\bigg(\frac{\partial L_1}{\partial \dot{\theta_1}}\bigg)-\frac{\partial L_1}{\partial \theta_1}+\frac{\partial \mathcal{F}}{\partial \dot{\theta_1}}= Q_1^{drive}
\end{eqnarray}
where $Q_1^{drive}$ is the driving force in the $\theta_1$ direction.
This gives us the equation
\begin{eqnarray}
\frac{m_1\ell^2\ddot{\theta_1}}{4}-\frac{m_1g\ell\sin\theta_1}{2}+k_1\theta_1\big(1+\alpha_1\theta_{1}^2\big)+\frac{b\ell^2{\dot{\theta_1}}}{4}= Q_1^{drive}
\end{eqnarray}
which is the equation of motion for one beam system.

Now we add a driving force to the system. We consider the driving force due to the wind and hence two possibilities. The first one is that of wind changing directions continuously, a situation typical of stormy conditions, and the other possibility is that of wind coming from a fixed horizontal direction accompanied by the modulations. The first force would be better modelled by a term $\ell f\cos \omega t/2$ and the second can 
be expressed mathematically as $\ell(d+f\cos \omega t) \cos \theta_1/2$ where $d$ is the average force in a given direction and we have multiplied by $\cos \theta_1$ as we need the component in the $\theta_1$ direction. 
On simplification and substituting $\ell=1$ and $m_1=1$, we get,
\begin{equation}\label{eq:d} \ddot{\theta} + b\dot{\theta} - 2g\sin{\theta} + 4k\theta (1+\alpha \theta^2) = 2 \left\{ \begin{array}{l} f\cos \omega t \\ (d+f\cos \omega t) \cos \theta \end{array} \right. \end{equation}
where the subscript of $\theta$ has been omitted.

To the best of our knowledge, there is no other system where such an equation has arisen in which both these nonlinear forces, the gravitational term as in pendulum and the cubic nonlinearity in the restoring force, are present. As a result no mathematical analysis of such an equation exist in the literature.
While the equations with the presence of these nonlinear forces separately have been solved in terms of Jacobi elliptic functions, the above equation with both the terms present doesn't seem to be amenable to analytic treatment. Even the convergence with an approximate method using Adomian decomposition \cite{AG} is very slow. \\

In \cite{LM}, the first two terms in the power series exapansion of sine were used leading to the equation:
\begin{equation}\label{eq:LM} \ddot{\theta} + b\dot{\theta} + (4k-2g)\theta + (4k\alpha+\frac{g}{3}) \theta^3  = 2f\cos(\omega t) \end{equation}
This is a modified Duffing's oscillator which has been studied extensively as an example of the simplest nonlinear generalization of
driven damped simple harmonic oscillator. The work by Miller~\cite{LM} was, to the best of our knowledge, the first instance of incorporating nonlinearity in the modeling of swaying of trees. Such an approximation would be clearly of use at low wind speeds. In this work, the effect of nonlinearity in the resonance curve was explored in detail. In the present work, we plan to study the chaotic properties of the solutions without making such an approximation.

\subsection{The case of two elements}
Now let us consider two segments, The Lagrangian takes the form:
\begin{eqnarray}
L_2&=&\frac{1}{2}l^2\bigg(\frac{m_1}{4}+m_2\bigg)\dot{\theta_1}^2+\frac{1}{2}l^2\bigg(\frac{m_2}{4}\bigg)\dot{\theta_2}^2+l^2\dot{\theta_1}\dot{\theta_2}\cos(\theta_2-\theta_1)\bigg(\frac{m_2}{2}\bigg)\nonumber\\
&&\;\;-gl\bigg(\frac{m_1}{2}+m_2\bigg)\cos\theta_1-gl\bigg(\frac{m_2}{2}\bigg)\cos\theta_2-\frac{1}{2}k_1\theta_1^2\nonumber\\
&&\;\;\;\;\;-\frac{1}{4}k_1\alpha_1\theta_1^4-\frac{1}{2}k_2(\theta_2-\theta_1)^2-\frac{1}{4}k_2\alpha_2(\theta_2-\theta_1)^4
\end{eqnarray}
Again the dissipation is incorporated through Rayleigh's dissipation function, which is given by 
\begin{eqnarray}
\mathcal{F}_2&=&\frac{1}{2}\bigg(b_x{v_{x1}}^2+b_y{v_{y1}}^2+b_x{v_{x2}}^2+b_y{v_{y2}}^2\bigg) \nonumber \\
&=&\frac{1}{2}\Bigg(b_x\bigg(\big({\frac{\ell}{2}\cos\theta_1\dot{\theta_1}}\big)^2+\big(\frac{\ell}{2}\cos\theta_2\dot{\theta_2}+\ell\cos\theta_1\dot{\theta_1}\big)^2\bigg)\nonumber\\
&&\;\;\;\;+b_y\bigg(\big(-\frac{\ell}{2}\sin\theta_1\dot{\theta_1}\big)^2+\big(-\frac{\ell}{2}\sin\theta_2\dot{\theta_2}-\ell\sin\theta_1\dot{\theta_1}\big)^2\bigg)\Bigg),\nonumber\\
\end{eqnarray}
and with the assumption that the dissipation along each direction is the  same, i.e. $b_x=b_y=b$, it simplifies to
\begin{eqnarray}
\mathcal{F}_2&=&\frac{1}{2}b\ell^2\Bigg(\frac{5{\dot{\theta_1}}^2}{4}+\frac{{\dot{\theta_2}}^2}{4}+\dot{\theta_1}\dot{\theta_2}\cos(\theta_2-\theta_1)\Bigg)
\end{eqnarray}

The Lagrange's equation of motion for a two element system becomes
\begin{eqnarray}
\frac{d}{dt}\bigg(\frac{\partial L_2}{\partial \dot{\theta_1}}\bigg)-\frac{\partial L_2}{\partial \theta_1}+\frac{\partial \mathcal{F}_2}{\partial \dot{\theta_1}}=Q_1^{drive}\\
\frac{d}{dt}\bigg(\frac{\partial L_2}{\partial \dot{\theta_2}}\bigg)-\frac{\partial L_2}{\partial \theta_2}+\frac{\partial \mathcal{F}_2}{\partial \dot{\theta_2}}=Q_2^{drive}
\end{eqnarray}
This gives us the equation
\begin{eqnarray}
\ell^2\bigg(\frac{m_1}{4}+m_2\bigg)\ddot{\theta_1}+\ell^2\bigg(\frac{m_2}{2}\bigg)\bigg(\ddot{\theta_2}\cos(\theta_2-\theta_1)-{\dot{\theta_2}}^2\sin(\theta_2-\theta_1)\bigg)-g\ell\bigg(\frac{m_1}{2}+m_2\bigg)\sin\theta_1\nonumber\\
\;\;+k_1\theta_1\big(1+\alpha_1\theta_{1}^2\big)-k_2(\theta_2-\theta_1)\bigg(1+\alpha_2(\theta_2-\theta_1)^2\bigg)+\frac{1}{2}b\ell^2\Bigg(\frac{5{\dot{\theta_1}}}{2}+\dot{\theta_2}\cos(\theta_2-\theta_1)\Bigg)&=&Q_1^{drive}\\
\ell^2\bigg(\frac{m_2}{4}\bigg)\ddot{\theta_2}+\ell^2\bigg(\frac{m_2}{2}\bigg)\bigg(\ddot{\theta_1}\cos(\theta_2-\theta_1)-{\dot{\theta_1}}^2\sin(\theta_2-\theta_1)\bigg)-g\ell\bigg(\frac{m_2}{2}\bigg)\sin\theta_2\nonumber\\
\;\;+k_2(\theta_2-\theta_1)\bigg(1+\alpha_2(\theta_2-\theta_1)^2\bigg)+\frac{1}{2}b\ell^2\Bigg(\frac{{\dot{\theta_2}}}{2}+\dot{\theta_1}\cos(\theta_2-\theta_1)\Bigg)&=&Q_2^{drive}\\\nonumber
\end{eqnarray}
which is the equation of motion for two beam system.
On simplication and rearranging coefficients, it becomes
\begin{eqnarray}
\ddot{\theta_1}&=&-\frac{2}{5}\ddot{\theta_2}\cos(\theta_2-\theta_1)+\frac{2}{5}{\dot{\theta_2}}^2\sin(\theta_2-\theta_1)+\frac{6}{5}g\sin\theta_1-\frac{4}{5}k_1\theta_1\big(1+\alpha_1\theta_{1}^2\big)\nonumber\\
&&\;\;+\frac{4}{5}k_2(\theta_2-\theta_1)\bigg(1+\alpha_2(\theta_2-\theta_1)^2\bigg)-\frac{2}{5}b\Bigg(\frac{5{\dot{\theta_1}}}{2}+\dot{\theta_2}\cos(\theta_2-\theta_1)\Bigg)+\frac{4}{5}Q_2^{drive}\nonumber\\
\ddot{\theta_2}&=&-2\ddot{\theta_1}\cos(\theta_2-\theta_1)-2{\dot{\theta_1}}^2\sin(\theta_2-\theta_1)+2g\sin\theta_2-4k_2(\theta_2-\theta_1)\bigg(1+\alpha_2(\theta_2-\theta_1)^2\bigg)\nonumber\\
&&\;\;-2b\Bigg(\frac{{\dot{\theta_2}}}{2}+\dot{\theta_1}\cos(\theta_2-\theta_1)\Bigg)+4Q_2^{drive}\nonumber
\end{eqnarray}
We eliminate the second derivatives on the RHS to obtain
\begin{eqnarray}\label{eq:th1}
\ddot{\theta_1}&=&\Bigg(\frac{1}{1-\frac{4}{5}{\cos(\theta_2-\theta_1)}^2}\Bigg)\Bigg(\frac{4}{5}{\dot{\theta_1}}^2\sin(\theta_2-\theta_1)\cos(\theta_2-\theta_1)-\frac{4}{5}g\sin\theta_2\cos(\theta_2-\theta_1)\nonumber\\
&&\;\;+\frac{8}{5}k_2(\theta_2-\theta_1)\cos(\theta_2-\theta_1)\bigg(1+\alpha_2(\theta_2-\theta_1)^2\bigg)+\frac{4}{5}b\dot{\theta_1}{{\cos(\theta_2-\theta_1)}^2}-\frac{8}{5}Q_2^{drive}\cos(\theta_2-\theta_1)\nonumber\\
&&\;\;+\frac{2}{5}{\dot{\theta_2}}^2\sin(\theta_2-\theta_1)+\frac{6}{5}g\sin\theta_1-\frac{4}{5}k_1\theta_1\big(1+\alpha_1\theta_{1}^2\big)+\frac{4}{5}k_2(\theta_2-\theta_1)\bigg(1+\alpha_2(\theta_2-\theta_1)^2\bigg)\nonumber\\
&&\;\;-b\dot{\theta_1}+\frac{4}{5}Q_2^{drive}\Bigg)\\
\ddot{\theta_2}&=&\Bigg(\frac{1}{1-\frac{4}{5}{\cos(\theta_2-\theta_1)}^2}\Bigg)\Bigg(-\frac{4}{5}{\dot{\theta_2}}^2\sin(\theta_2-\theta_1)\cos(\theta_2-\theta_1)-\frac{12}{5}g\sin\theta_1\cos(\theta_2-\theta_1)\nonumber\\
&&\;\;+\frac{8}{5}k_1\theta_1\cos(\theta_2-\theta_1)\bigg(1+\alpha_1\theta_1^2\bigg)-\frac{8}{5}k_2(\theta_2-\theta_1)\cos(\theta_2-\theta_1)\bigg(1+\alpha_2(\theta_2-\theta_1)^2\bigg)\nonumber\\
&&\;\;+\frac{4}{5}b\dot{\theta_2}{{\cos(\theta_2-\theta_1)}^2}-\frac{8}{5}Q_2^{drive}\cos(\theta_2-\theta_1)-2{\dot{\theta_1}}^2\sin(\theta_2-\theta_1)+2g\sin\theta_2\nonumber\\
&&\;\;-4k_2(\theta_2-\theta_1)\bigg(1+\alpha_2(\theta_2-\theta_1)^2\bigg)-b\dot{\theta_2}+4Q_2^{drive}\Bigg)\label{eq:th2}
\end{eqnarray}

\section{Results}\label{se:results}

The primary aim of this work is to obtain some insight into the dynamics of this model with single element as it forms the building block of the general model.
We have carried out extensive numerical simulations to understand the behaviour of this nonlinear single element chimney model described by the equation~(\ref{eq:d}). This understanding would then be useful later for the model with several elements. Firstly, we are interested in understanding the chaotic solutions of this model and hence the values of the largest Lyapunov exponent for various values of the parameters. For comparison we also study the chaotic properties of two element system. We would also like to study the effect of directionality in the wind on its chaotic properties for different parameter values.

\subsection{Riddled basins of attraction}

\begin{figure}[h]
         \includegraphics[width=8cm,scale=1]{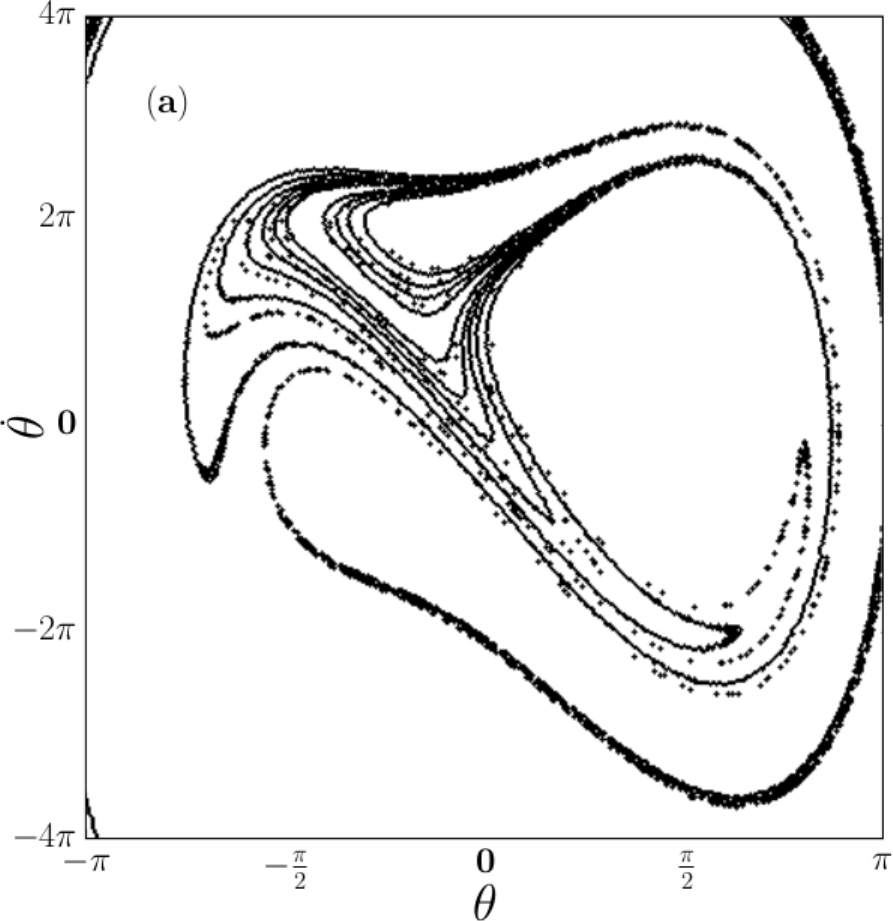}%
\includegraphics[width=8cm,scale=1]{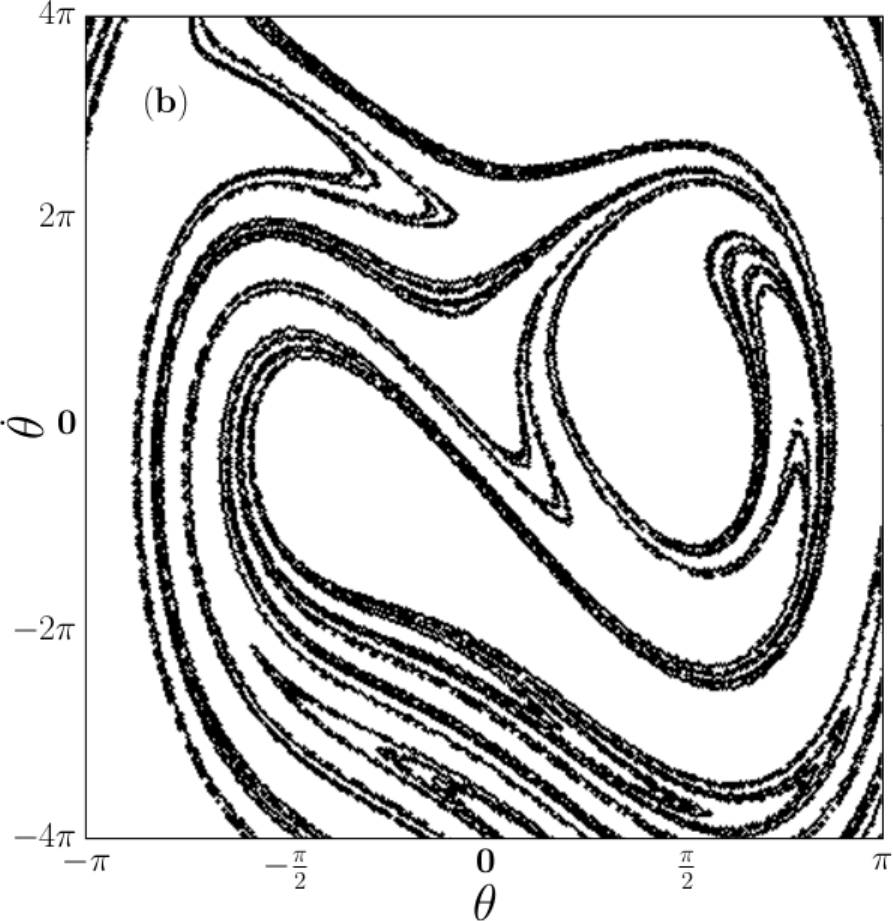}\\
\includegraphics[width=8cm,scale=1]{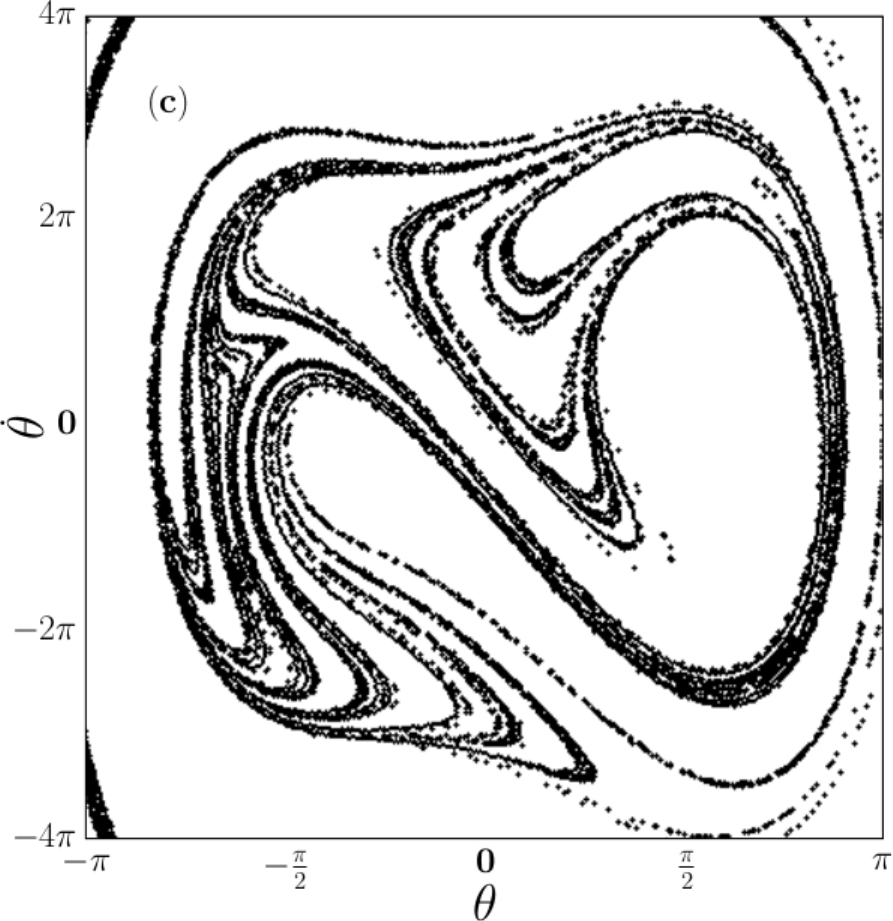}%
\includegraphics[width=8cm,scale=1]{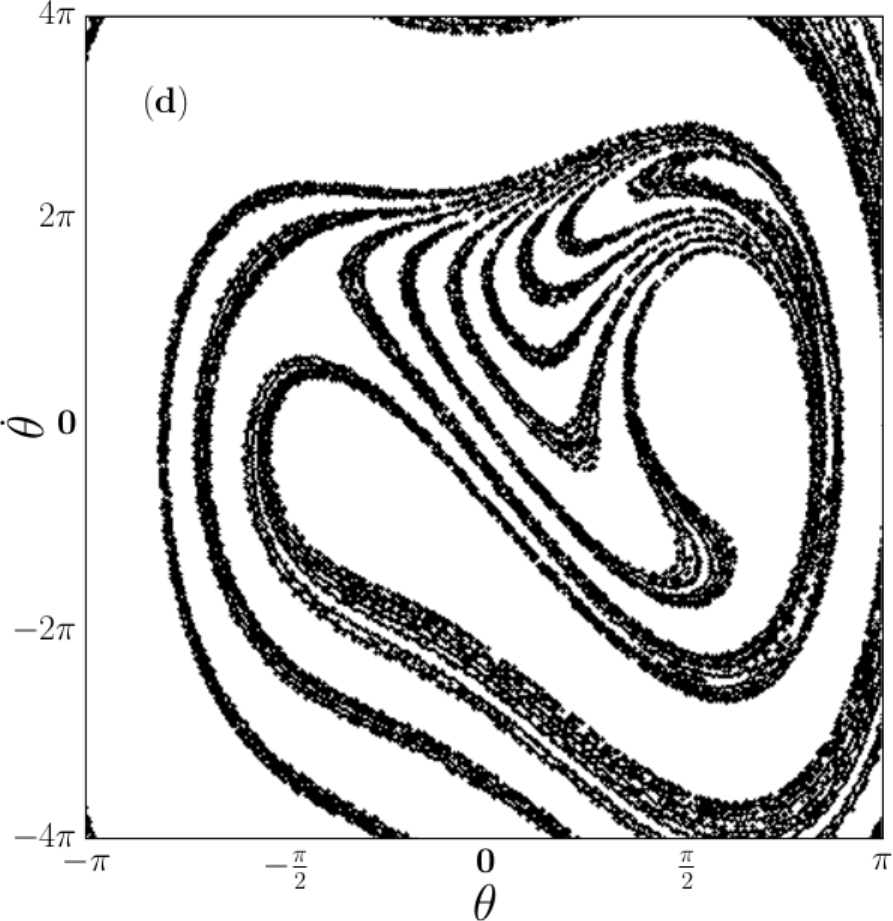}%
       \caption{Fractal basin boundaries for $k=1.0, \alpha=1.0, b=1.0$ (a) $f=2.0, \omega = 1.0, D_f=1.46$, (b) $f=2.0, \omega=1.8, D_f=1.61$, (c) $f=2.2, \omega=0.8, D_f=1.59$, (d) $f=2.2, \omega=1.2, D_f=1.64.$}\label{fig:bb}
  \end{figure}

The Duffing's oscillator has double well potential for all negative values of $k$ and $\alpha$ but our system, owing to the gravitational term, has double well potential even for small positive values of $k$ ($<g/2$) when $\alpha$ is also positive. Here, we restrict ourselves to the range of parameters leading to the double well potential. That is, there are two stable fixed points. In the case of Duffing's oscillator it is known that when the motion is regular the basins of attraction of these two fixed points have smooth boundary but as the values of $\omega$ and $f$ are increased the motion becomes irregular and the basin boundary starts to intersect with each other leading to a fractal nature.

We study the basins of attraction of the stable points and find that for sufficiently large values of $f$ and $\omega$ the basins become intertwined and the basin boundaries become fractal. Fig.~\ref{fig:bb} shows some examples. We find that the fractal dimensions $D_f$ lie around 1.5. This is similar to other systems reportd in the literature~\cite{ML, GOY}. 
  
\begin{figure}
\includegraphics[width=0.5\linewidth]{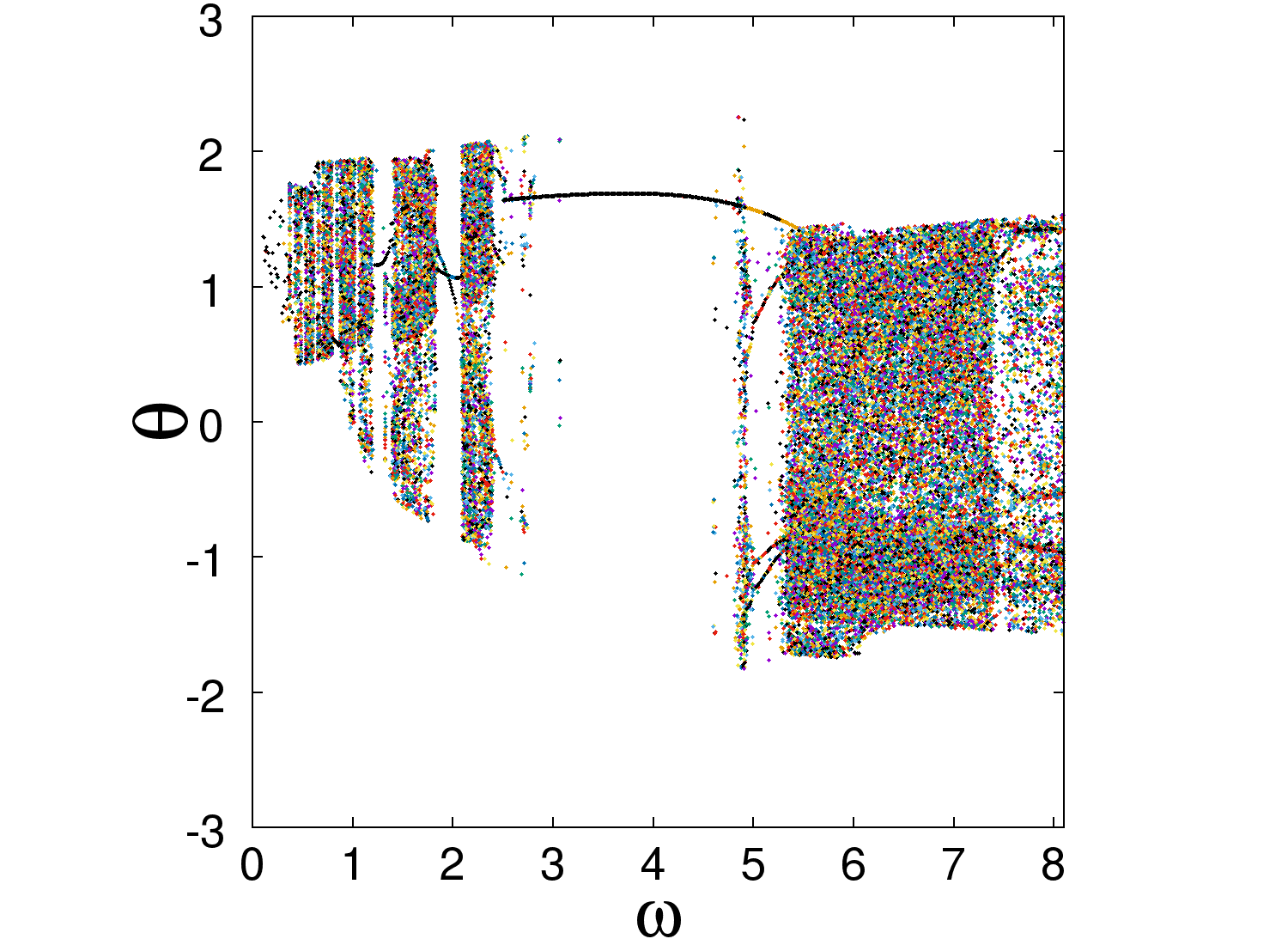}
    \caption{Poincare Section for $k=2.0, \alpha=1.0, b=0.5, f=4.4$}\label{fig:poincare}
\end{figure}
\begin{figure}
\includegraphics[width=0.5\linewidth]{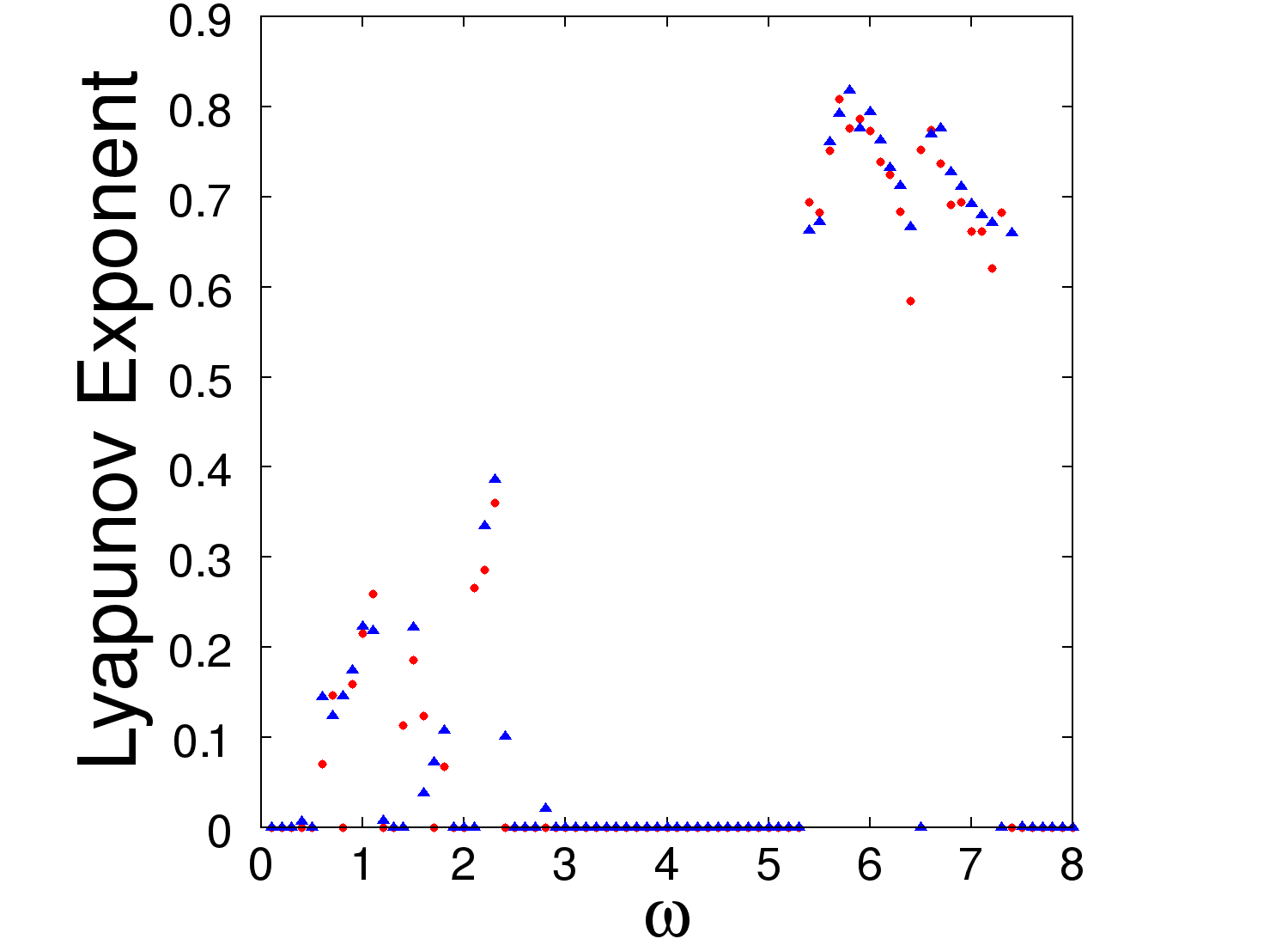}
  \caption{ Lyapunov Exponent for $k=2.0, \alpha=1.0, b=0.5, f=4.4$. The filled circles are for the equation with full sine term (Eq.~(\ref{eq:d})) and filled triangles represent truncated system~(Eq.~(\ref{eq:LM})). }\label{fig:lyap}
\end{figure}

\subsection{Lyapunov exponents}
\subsubsection{The case of a single element}
Then, we compute the largest Lyapunov exponents for various values of parameters
in order to check for the existence of chaotic motion. The largest Lyapunov exponent can be estimated by two different approaches: (i) by generating the divergence in the trajectory directly from the governing equation and thier Jacobians \cite{WSSV} and (ii) by generating a time series from the solution of the differential equation and then using softwares like TISEAN \cite{RCL}  or TSTOOL \cite{PU}. We have tried all these ways, though here we report the results of the first approach. It is worthwhile to note that all the approaches lead to consistent conclusion about the existence of chaos though the exact positive values of the largest Lyapunov exponents differed from method to method.
 We observe positive largest Lyapunov exponents for wide parameter ranges implying the chaotic motion.
In Fig.~\ref{fig:poincare}, we see the Poincare section for the system with full sine term for some values of parameters ($k=2.0, \alpha=1.0, b=0.5, f=4.4$). It shows the ranges of values of $\omega$s for which the motion seems irregular. We find, as shown in Fig.~\ref{fig:lyap}, that in several of these ranges of $\omega$ values the largest Lyapunov exponent is positive. It is interesting to note that the Poincare section for the truncated system is also almost identical to that shown in Fig.~\ref{fig:poincare} but the values of the Lyapunov exponents in the choatic region are generally different. Fig.~\ref{fig:lyap} also depicts the Lyapunov exponents for the system with truncated sine of Eq.~(\ref{eq:LM}). We find that there is a considerable difference in the Lyapunov exponent of the two systems for smaller range of $\omega$ values ($\omega < 3$) but not so in the higher range ($\omega > 5$). Incidently, this higher range of $\omega$ values where we see chaotic solutions corresponds to the range around the resonance and the lower range of $\omega$ with positive Lyapunov exponents corresponds to the subharmonic resonances.

\subsubsection{The two element case}

\begin{figure}
	\includegraphics[width=0.5\linewidth,scale=1.0]{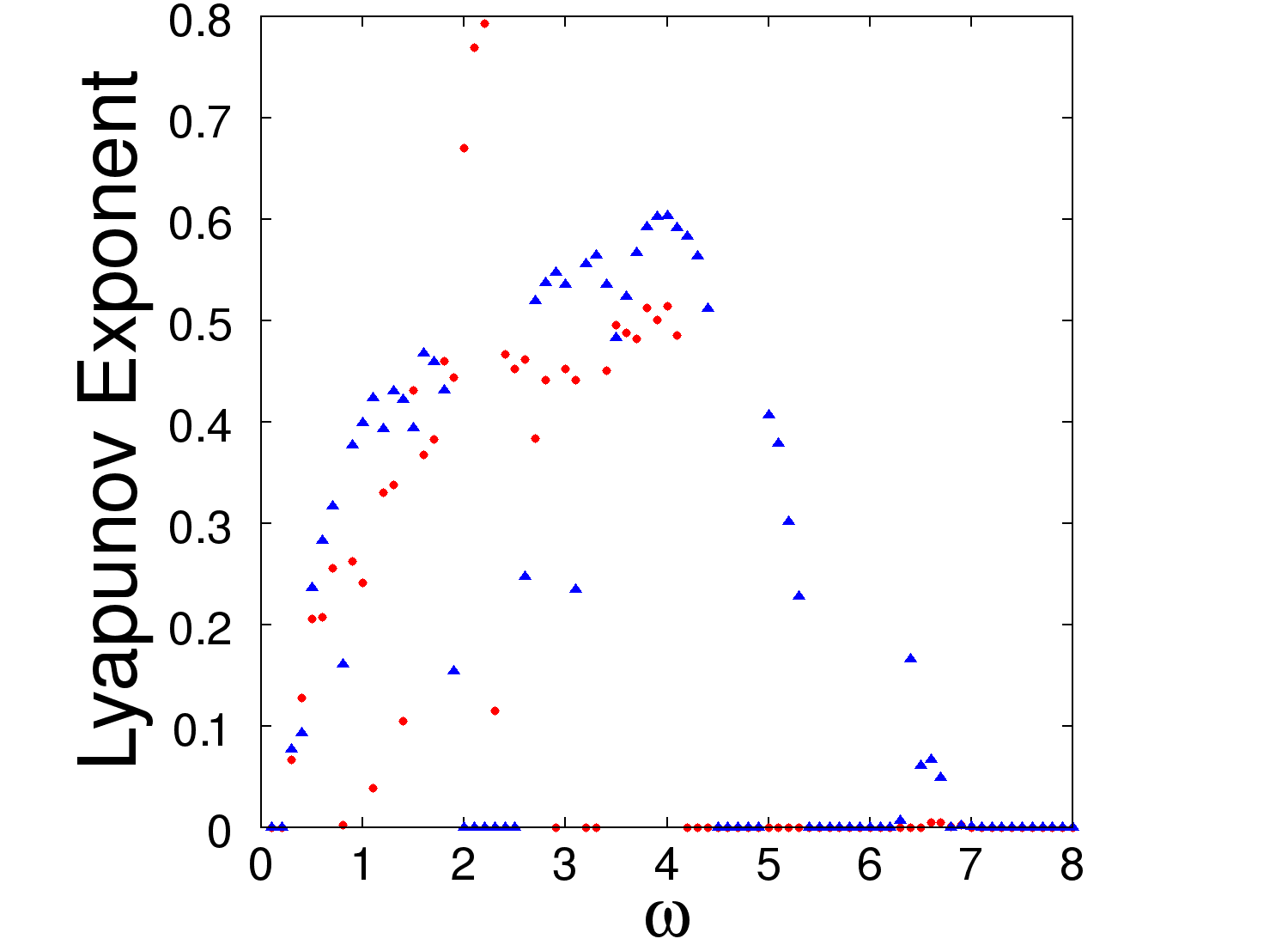}
	\caption{Lyapunov Exponent for two element system for $k_1=1, k_2=1, \alpha_1=1, \alpha_2=1, b=0.5, f=11.6$. The filled circles are for the equations (\ref{eq:th1}) and (\ref{eq:th2}) with full sine term and filled triangles are for the same system except that $g\sin(\theta_i)$ is replaced by $g(\theta_i-{\theta_i}^3/6)$}
\label{fig:lyap2beam}
	\end{figure}

In order to better understand the difference between the effect of full sine gravitational term and its truncation to cubic order we find the Lyapunov exponent of the two beam system. The Fig.~\ref{fig:lyap2beam} depicts the results. We observe that the values of the Lyapunov exponents are quite different especially as compared to the single beam case. Moreover, there are some values of $\omega$ for which the truncated system is chaotic but the system without approximation isn't.

\subsection{Effect of directionality}
As discussed in~\cite{HKW}, considering the effect of directionality in the wind is important especially for large wind speeds.
In this section, we study the existence of chaos as the parameter $d$ in Eq.~(\ref{eq:d}) is varied. This parameter adds a DC shift to the otherwise periodically varying force. So a non-zero $d$ means that the wind is flowing in certain direction modulated by periodic variations. As remarked before, such a wind is usually horizontal and hence only the component perpendicular to the segment will lead to the angular displacement. This makes it necessary to multiply the driving force by $\cos \theta$. This introduces a $\theta$ dependence on the right-hand-side of the equation. We observe that this multiplication by $\cos \theta$ leads, in general, to suppression of chaos. That is, for example, it is seen that the bands of chaotic behaviour in Fig.~\ref{fig:poincare} become smaller when the other parameters are kept the same. 

\begin{figure}
	\includegraphics[width=0.33\linewidth,scale=0.5]{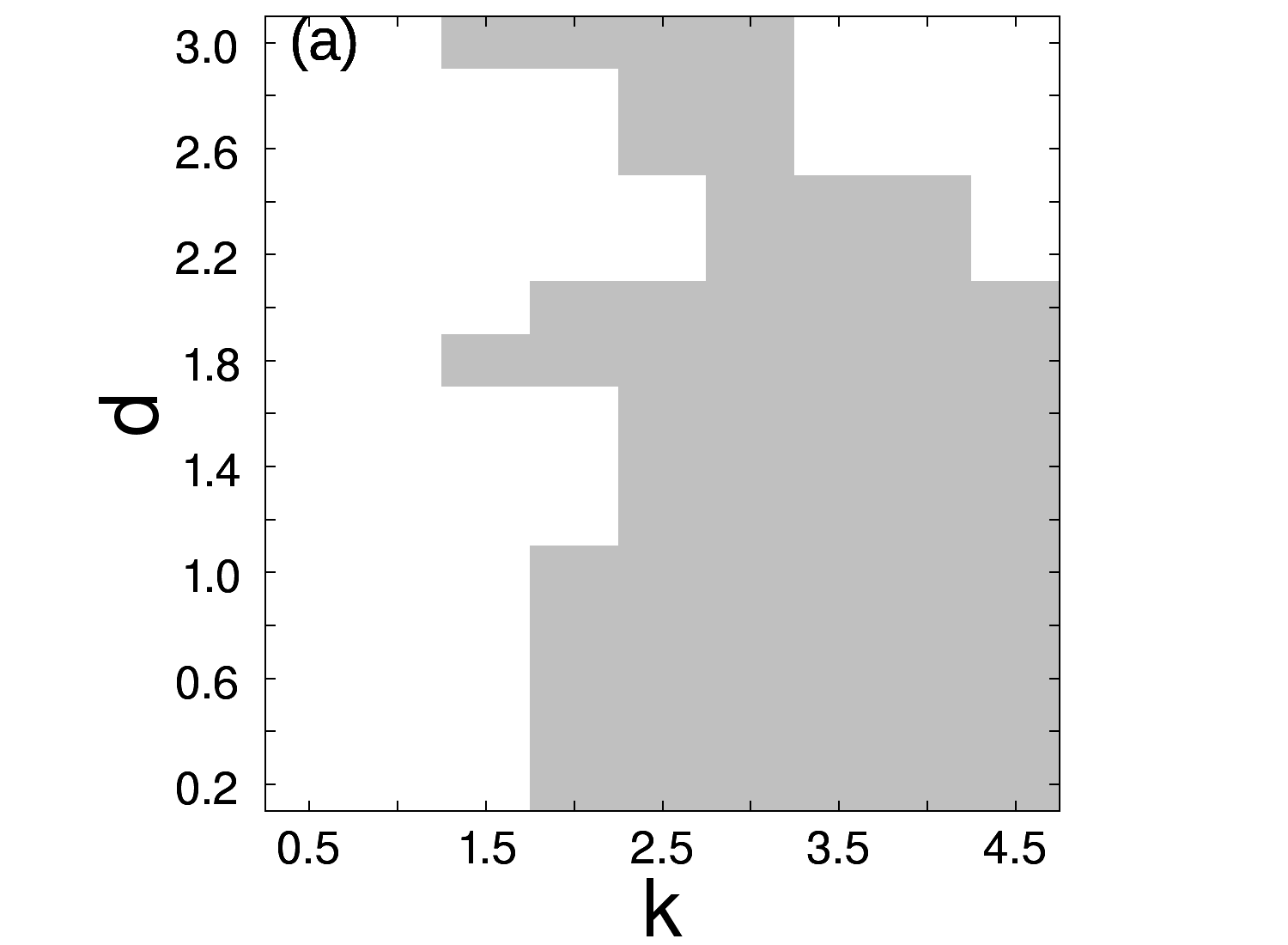}
        \includegraphics[width=0.33\linewidth,scale=0.5]{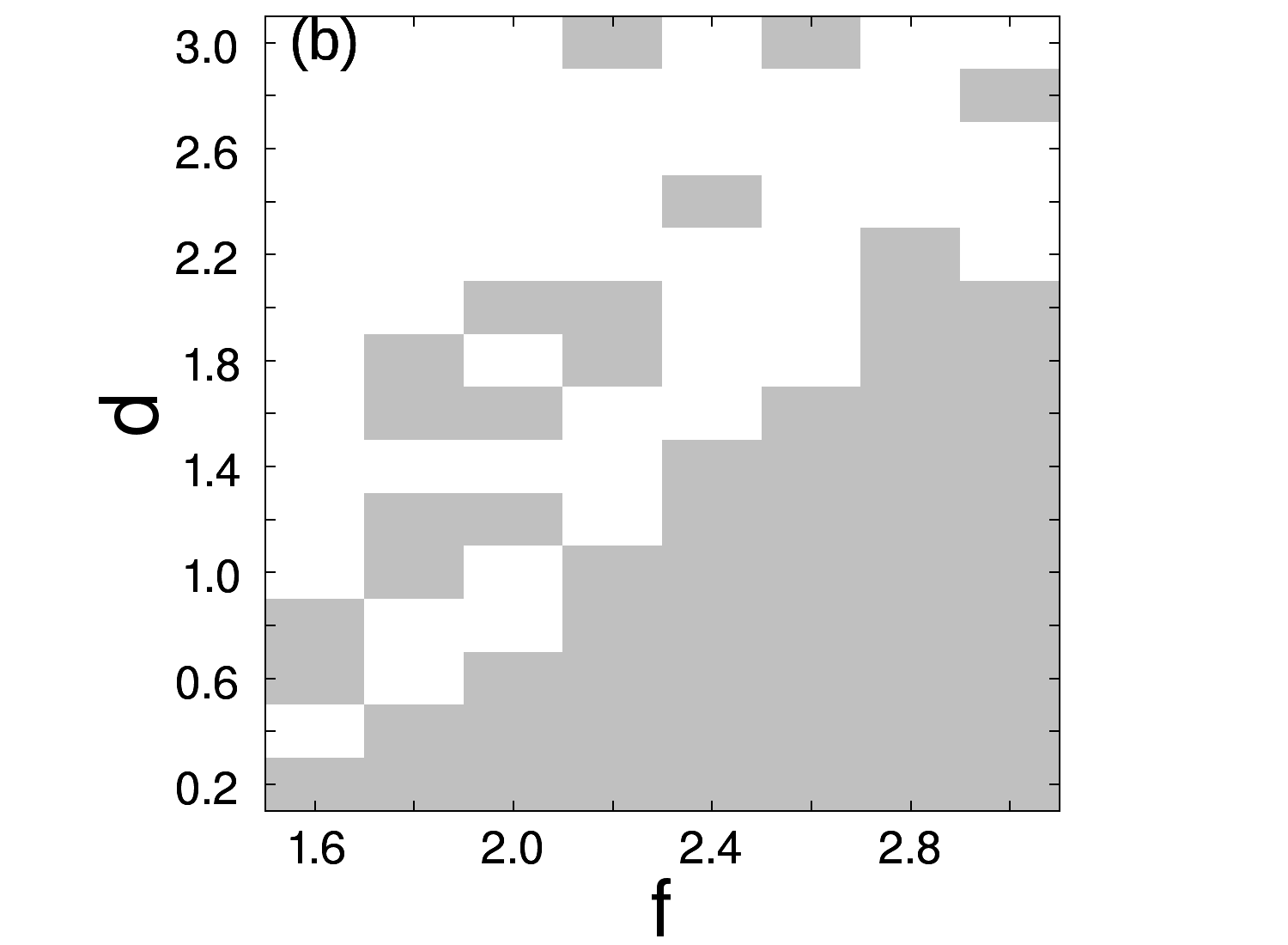}
        \includegraphics[width=0.33\linewidth,scale=0.5]{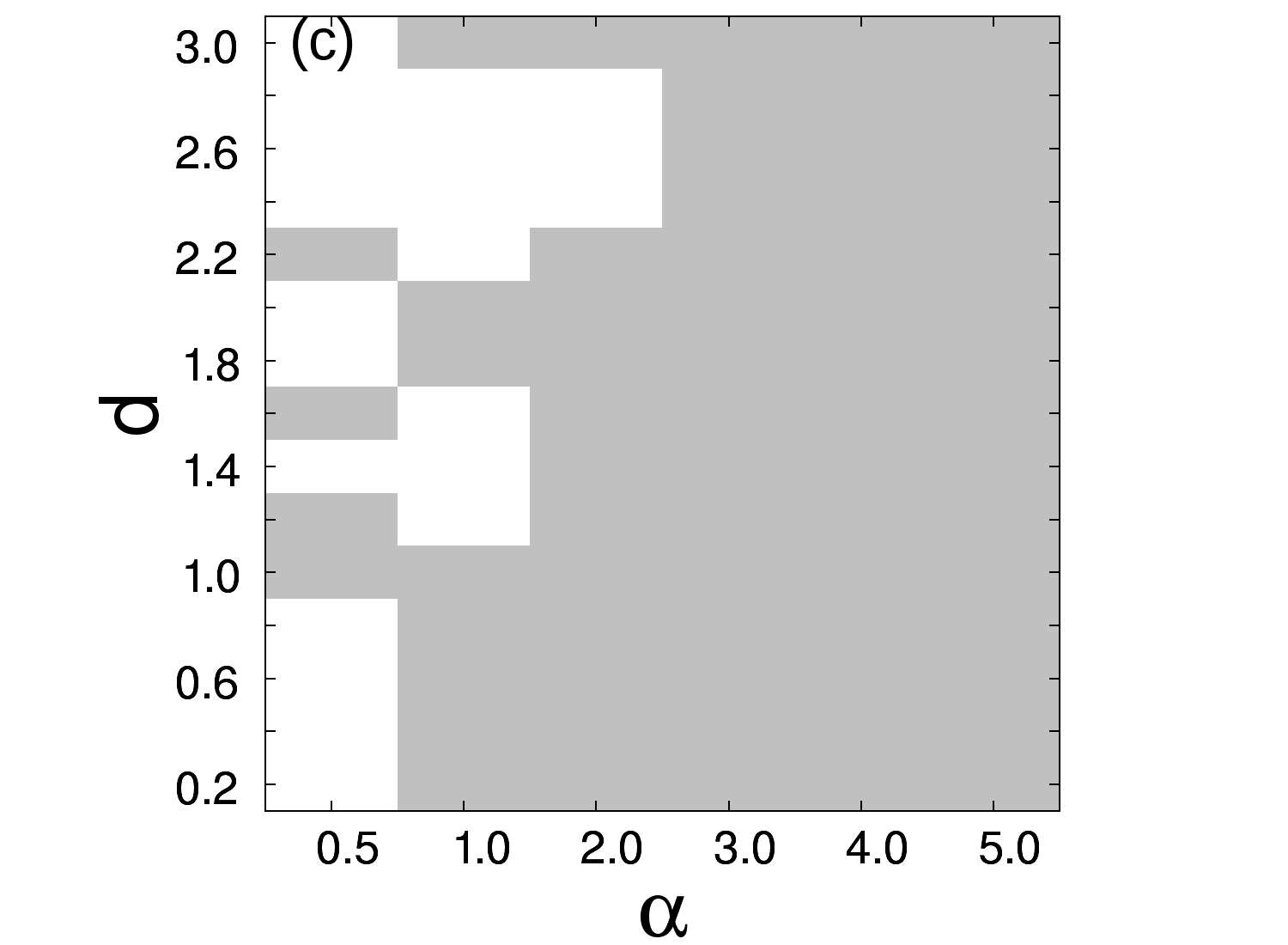}
	\caption{ White region implies that no chaos was observed for given values of parameters (in (a) $k$ and $d$, in (b) $f$ and $d$ and in (c) $\alpha$ and $d$) for the values of $\omega$ between 0 and 8 whereas the grey region corresponds to the existence of chaos for some values of $\omega$ in this range. }\label{fig:dcos}
	\end{figure}

We now further study the effect of varying $d$ and its dependence on other parameters, $k$, $f$ and $\alpha$.
 In general we find that the chaos is further suppressed as $d$ is increased keeping $f$ and $\alpha$ fixed. The Fig.~\ref{fig:dcos}a shows the effect of varying $d$ for different values of $k$. The white region corresponds to no evidence of positive Lyapunov exponent for the range of $\omega$ (between 0 and 8) values explored. Whereas, the grey region corresponds to existence of chaos atleast for some values of $\omega$. Interestingly, the suppression of chaos with increasing $d$ is more prominent at larger values of $k$. This is surprising because, at smaller values of $d$, it is for this range of $k$ that one observes more robust chaos, in the sense that the system is chaotic for larger range of $\omega$ values and the values of Lyapunov exponents are relatively larger. The result of changing $d$ and $f$ keeping $k$ and $\alpha$ fixed is shown in the Fig.~\ref{fig:dcos}b. Here too we see that the chaos disappears for larger value of $d$. However, as expected, the range of values of $d$ over which chaos exists 
increases with $f$. Finally, in the Fig.~\ref{fig:dcos}c, we show the results when $d$ and $\alpha$ is varied
keeping $k$ and $f$ fixed. Here we do not see this feature of suppression of chaos as $d$ is increased at least for the range of parameters studied. In fact, there seems to be a critical value of $\alpha$ above which one observes chaos even for larger values of $d$. 

\section{Conclusion}\label{se:conclud}
We have begun a complete nonlinear analysis of swaying of trees. Such studies are of interest to forest scientistits interested in minimizing the loss of wood in, say, stormy conditions and also to computer scientists interested in building realistic animation of moving trees or jungles. Though, it is known that the biological materials show nonlinear stiffness properties, the models built by computer scientists are exclusively based on linear restoring forces whereas the studies stemming from the plant biologists have only started to include some nonlinear properties. We have planned to carry out a full-fledged study of the swaying of trees incorporating the nonlinearity as much as possible. As a starting point, we have considered the chimney model which was used before for the same purpose but without incorporating nonlinearity. It consists of several segments connected end to end and erected from the ground. This choice of the model would easily allow to add the branches later. There is a  restoring force between the joints and also at the base of the first element and the ground. There is also the gravitational force acting on each of these elements. We have derived general equations of motion by reformulating this model using Lagrangian formulation by taking the cubic nonlinearity in the restoring force and the 
full sine term for the gravitational force.

Here our attention is primarily on the single element model but have considered the two element case too. We have analyzed various nonlinear dynamical aspects without any consideration to the biological values of the parameters.   We found that there exist positive Lyapunov exponent in a certain region of parameter space. The Lyapunov exponents for the system with truncated system are not generally the same as compared with the system with full sine term. The sine term in the gravitational force introduces a length scale in the problem which can lead to qualitatively different behavior with branched structure and at high wind speeds. In fact, we find that, in the two element case, there are values of parameters for which the truncated system is chaotic but there is no chaos for the system with full sine term. We have also studied the effect of the directionality in the wind on the nature of chaos and found that the chaos gets suppressed as the wind velocity increases in certain direction.

In future, it is planned to study the model further with multiple segments and also branched structures. The branch structures could consist of simple structure with few branches or a selfsimilar structure with several subbranches. Also, it would be of interest to study the effect of different driving forces. For a complete understanding, the inclusion of torsional oscillations would also be worthwhile.

It is also planned to carry out the comparison with experimental data. This will be done with the data already available in the literature and also on the data specially obtained by measurements on the video recordings of small plants and grass-like plants.

\nonumsection{Acknowledgments} \noindent 
KMK and ARVK would like to thank Science and Engineering Research Board (SERB), India for financial assistance during this work.
We also thank M. R. Press and Sanjay Sane for carefully reading the manuscript.


\begin{thebibliography}{9} 
\bibitem[Adomian(1991)]{AG} Adomian G. [1991] ``A review of the decomposition method and some recent results for nonlinear equations," {\it Computers Math. Applic.} {\bf 21}, 101-127.
\bibitem[Akagi {\it et al.}(2006)]{AK} Akagi Y. and Kitajima K. [2006] ``Computer animation of swaying trees based on physical simulation," {\it Computers and Graphics} {\bf 30}, 529-539.
\bibitem[Barbacci {\it et al.}(2013)]{BDH} Barbacci A., Diener J., Hemon P., Adam B., Dones N., Reveret L. and Moulia B. [2013] ``A robust videogrametric method for the velocimetry of wind induced motion in trees," {\it Agricultural and Forest Meteorology} {\bf  184}, 220-229.
\bibitem[De Langre(2008)]{Lan} De Langre E. [2008] ``Effects of Wind on Plants," {\it Annu. Rev. Fluid Mech.} {\bf 40}, 141-168.
\bibitem[Diener {\it et al.}(2009)]{DRBR} Diener J., Rodriguez M., Baboud L. and Reveret L. [2009] ``Wind projection basis for real-time animation of trees," {\it Computer Graphics Forum (Proceedings Eurographics)} {\bf 28}, 533-540.
\bibitem[Grebogi {\it et al.}(1987)]{GOY} Grebogi C., Ott E. and Yorke J. A. [1987] ``Chaos, Strange Attractors, and Fractal Basin Boundaries in Nonlinear Dynamics," {\it Science} {\bf 238}, 632-638.
\bibitem[Habel {\it et al.}(2009)]{HKW} Habel R., Kusternig A. and Wimmer M. [2009] ``Physically guided animation of trees," {\it Computer Graphics Forum (Proceedings Eurographics)} {\bf 28}, 523-533.
\bibitem[Hassinen {\it et al.}(1998)]{HLP} Hassinen A., Lemettinen M., Peltola H., Kellomaki S. and Gardiner B. [1998] ``A prism-based system for monitoring the swaying of trees under wind loading," {\it Agricultural and Forest Meteorology} {\bf 90}, 187-194.
\bibitem[Holmes(1979)]{HP} Holmes P.[1979] ``A Nonlinear Oscillator with a Strange Attractor," {\it Phil. Transc. of the Royal Soci. of Lon.} {\bf 292}, 419-448.

\bibitem[Hu {\it et al.}(2017)]{HZXI} Hu S., Zhang Z., Xie H. and Igarashi T. [2017] ``Data-driven modeling and animation of outdoor trees through interactive apporach," {\it The Visual Computer} {\bf 33}, 1017-1027.
\bibitem[Kerzenmacher \& Gardiner(1998)]{KG} Kerzenmacher T. and Gardiner B. [1998] ``A mathematical model to describe the dynamic response of a spruce tree to the wind," {\it Trees} {\bf 12}, 385-394.
\bibitem[Miller(2005)]{LM} Miller L. [2005] ``Structural dynamics and resonance in plants with nonlinear stiffness," {\it J. Theor. Biol.} {\bf 234}, 512-524.
\bibitem[Moon \& Li (1985)]{ML} Moon F. C. and Li G.-X. [1985] ``Fractal Basin Boundaries and Homoclinic Orbits for Periodic Motion in a Two Well Potential" {\it Phys. Rev. Lett} {\bf 55}, 1439-1442.
\bibitem[Moore \&  Maguire(2005)]{MM} Moore, J. R. and Maguire D. A. [2005] ``Natural sway frequencies and damping ratios of trees: influence of crown structure," {\it Trees} {\bf 19}, 363-373.
\bibitem[Murphy \& Rudnicki(2012)]{MR} Murphy K. D. and Rudnicki M. [2012] ``A Physics-based link model for tree vibrations," {\it Ameri. J. Bot.} {\bf 99}, 1918-1929.
\bibitem[Parlitz(1998)]{PU} Parlitz U. [1998] ``Nonlinear time series analysis," in: J.A.K. Suykens, J. Vandewalle (Eds.),  {\it Nonlinear Modeling},  209-239.
\bibitem[Ramasubramanian \&  Sriram(1999)]{RS} Ramasubramanian K. and Sriram M. S. [1999] ``Alternative algorithm for the computation of Lyapunov spectra of dynamical systems," {\it Phys. Rev. E} {\bf 60}, R1126. 
\bibitem[Rosenstein {\it et al.}(1993)]{RCL} Rosenstein M. T., Collins J. J. and De Luca C. J. [1993] ``A practical method for calculating largest Lyapunov exponents from small data sets," {\it Physica D} {\bf 65},  117-134.
\bibitem[Sellier {\it et al.}(2006)]{SFP} Sellier D., Fourcaud T. and Lac P. [2006] ``A finite element model for investigating effects of aerial architecture on tree oscillations." {\it Tree Physiology} {\bf 26}, 799-806.
\bibitem[Oliapuram {\it et al.}(2010)]{OK} Oliapuram N. J. and Kumar S. [2010] ``Realtime forest animation in wind," {\it  Proceedings  of  the  Seventh  Indian Conference on Computer Vision, Graphics and Image Processing, ICVGIP}{\bf 10}, 197-204.
\bibitem[Ota {\it et al.}(2003)]{OTFFMC} Ota S., Tamura M., Fujita K., Fujimoto T., Muraoka K. and Chiba N. [2003] ``$1/f^{\beta}$ Noise-based real-time animation of trees swaying in wind fields," {\it  Proceedings Computer Graphics International}, 52-59.

\bibitem[Theckes {\it et al.}(2011)]{TLB} Theckes B., de Langre E. and Boutillon X. [2011] ``Damping by branching: a bioinspiration from trees," {\it Bioinspir. and Biomim.}  {\bf 6}, 1-11. 
\bibitem[Vincent(1990)]{Vin} Vincent, J. [1990] ``Structural Biomaterials," {\it Princeton University Press, Princeton}.
\bibitem[Wolf {\it et al.}(1985)]{WSSV} Wolf A., Swift J. B., Swinney H. L., and Vastano J. A. [1985] ``Determining Lyapunov Exponents from a Time Series," {\it  Physica D} {\bf 16}, 285-317.


\end{thebibliography}
\end{document}